\documentclass[twocolumn]{aastex61}

\newcommand\aastex{AAS\TeX}

\received{}
\revised{}
\accepted{}

\shorttitle{\aastex\ Red supergiant stars in IC 1613}
\shortauthors{Chun et al.}

\begin{document}

\title{Red supergiant stars in IC 1613 and metallicity-dependent mixing length in
the evolutionary model}

\correspondingauthor{Sang-Hyun Chun}
\email{shyunc@kasi.re.kr}
\author[0000-0002-6154-7558]{Sang-Hyun Chun}
\affiliation{Department of Physics and Astronomy, Seoul National University, 08826, Seoul, Republic of Korea}
\affiliation{Korea Astronomy and Space Science Institute, 776 Daedeokdae-ro, Yuseong-gu, Daejeon 34055, Republic of Korea} 
\author{Sung-Chul Yoon}
\affiliation{Department of Physics and Astronomy, Seoul National University, 08826, Seoul, Republic of Korea}
\affiliation{SNU Astronomy Research Center, Seoul National University, Seoul 08826, Korea}
\author{Heeyoung Oh}
\affiliation{Korea Astronomy and Space Science Institute, 776 Daedeokdae-ro, Yuseong-gu, Daejeon 34055, Republic of Korea}
\author{Byeong-Gon Park}
\affiliation{Korea Astronomy and Space Science Institute, 776 Daedeokdae-ro, Yuseong-gu, Daejeon 34055, Republic of Korea}
\author{Narae Hwang}
\affiliation{Korea Astronomy and Space Science Institute, 776 Daedeokdae-ro, Yuseong-gu, Daejeon 34055, Republic of Korea}




\begin{abstract}
We report a spectroscopic study on red supergiant stars (RSGs) in the irregular
dwarf galaxy IC 1613 in the Local Group.  We derive the effective temperatures ($T_\mathrm{eff}$)
and metallicities of 14 RSGs by synthetic spectral fitting to the spectra
observed with the MMIRS instrument on the MMT telescope for a wavelength range
from 1.16 $\mu$m to 1.23 $\mu$m.  A weak bimodal distribution of the
RSG metallicity centered on the [Fe/H]=$-0.65$ is found, which is slightly lower than or
comparable to that of the Small Magellanic Cloud (SMC).  There is no evidence
for spatial segregation between the metal rich ([Fe/H]$>-0.65$)  and poor
([Fe/H]$<-0.65$) RSGs throughout the galaxy.  
The mean effective temperature of our RSG sample in IC 1613 is higher by about 250 K than
that of the SMC. However, no correlation between $T_\mathrm{eff}$
and metallicity within our RSG sample is found.  We calibrate
the convective mixing length ($\alpha_{\mathrm{MLT}}$) by comparing stellar evolutionary
tracks with the RSG positions on the HR diagram, finding that models with
$\alpha_{\mathrm{MLT}}=2.2-2.4 H_P$ can best reproduce the effective temperatures of the 
RSGs in IC 1613 for both Schwarzschild and Ledoux convection criteria. 
This result supports our previous study that
a metallicity dependent mixing length is needed
to explain the RSG temperatures observed in the Local Group, 
but we find that this dependency becomes relatively weak for RSGs
having a metallicity equal to or less than the SMC metallicity. 
\end{abstract}




\section{Introduction}

Red supergiants (RSGs) are massive stars with initial masses between about 9
$M_\sun$ and 30 $M_\sun$ at the post-main sequence evolutionary phase,  of which the
hydrogen envelopes are extended to several hundred solar radii
\citep[e.g.,][]{Levesque2005,Ekstrom2012}.  It is important to understand their
physical properties such as the effective temperature, radius, and mass-loss
rate in order to understand the evolutionary history of RSGs and the exact
nature of Type II supernova progenitors.  

The effective temperatures of RSGs are determined by the Hayashi
limit~\citep{Hayashi1961}.  Stellar evolution models predict that this limit
mainly depends on two factors for a given initial mass.  One is metallicity,
which greatly influences stellar opacity, and the other is the so-called mixing
length, which characterizes the energy transport efficiency by convection in
the convective hydrogen
envelope~\citep[e.g.,][]{Elias1985,Ekstrom2012,Chun2018}.  The Hayashi limit
shifts to a higher temperature for a lower metallicity and a larger mixing
length (hereafter, the mixing length is denoted by $\alpha_\mathrm{MLT}$ which
is given in units of the local pressure scale height $H_P$).  The mixing length
is a free parameter and needs an empirical calibration.  Most stellar evolution
models adopt a fixed mixing length of $\alpha_\mathrm{MLT} \approx 2.0 $
following the calibration result with the Sun.  However, some studies provide
evidence that a fixed mixing length is not suitable for all types of
stars~\citep[e.g.,][]{Guenther2000,Bonaca2012,Tayar2017} and that the mixing length depends on
metallicity for both low- and high-mass stars~\citep{Joyce2018, Chun2018}.  
More recently,~\citet{Gonzalez2021} also found that the standard solar mixing length is not
applicable to RSGs in their comparison of effective temperature of Wolf-Lundmark-Mellote (WLM) galaxy 
with the Geneva evolution models.

\citet{Chun2018} calibrated the mixing length values in RSGs for
various metallicities by comparing a grid of massive star evolution models with
RSGs observed in the nearby universe, and found that the mixing length
decreases for decreasing metallicity. Such a calibration of the mixing length
is also important for theoretical predictions of the structure (in particular,
the radius) of Type II supernova progenitors as discussed by~\citet{Chun2018}. 

The effective temperatures of RSGs can be inferred from their spectra 
~\citep{Levesque2005,Davies2013,Tabernero2018}.
However, spectroscopic studies on RSGs in the extragalactic galaxies are largely limited to the Magellanic
Clouds~\citep{Levesque2006}, M31~\citep{Massey2009,Gordon2016},
and M33~\citep{Drout2012,Gordon2016}.  Only a small number of RSGs (about $10$
RSGs) in irregular galaxies with lower metallicity environments than the LMC
metallicity have been investigated in several
studies~\citep{Levesque2012,Brita2014,Brita2015,Patrick2015,Garcia2018}.  
Therefore, a spectroscopic investigation of a larger sample
of RSGs in extragalactic galaxies with a wide range of metallicity from
[Fe/H]$=-0.49$ (e.g., SMC) to [Fe/H]$=-1.0$ (e.g., Sextans A and Sagittarius
Dwarf Irregular galaxy) would provide important information to constrain
stellar evolution models of RSGs.  

The photometric identification of RSGs for extragalactic galaxies~\citep[e.g., NGC 4449, NGC 5055, and NGC 5457 by][]{Chun2017} including the dwarf irregular galaxy (dIrr)
IC 1613~\citep{Chun2015} has been conducted.
IC 1613 is
an excellent laboratory to study  RSGs for a number of reasons. IC 1613 is a
gas-rich, isolated galaxy with no past interaction with other galaxies. The
distance of about 730 kpc~\citep{Dolphin2001,Piet2006}, a low inclination of
$i=38\arcdeg$~\citep{Lake1989} and the high Galactic latitude implying a low
extinction value of
$E(B-V)=0.02-0.04$~\citep{Sandage1971,Freedman1988a,Cole1999} provide a great
opportunity to access the resolved stellar populations with a reasonable
exposure time from a ground-based telescope. Several studies indicate that IC
1613 has a nearly constant star formation rate over its entire
lifetime~\citep{Cole1999,Skillman2003,Skillman2014}, hosting old ($>10$ Gyr),
intermediate-age ($1-10$ Gyr), and young ($<$1 Gyr) stellar populations.
However, most studies of the stellar populations in this galaxy have been
focused on the old and intermediate-age
stars having metallicity of [Fe/H]$=-1.75\sim-1.15$ ~\citep{Cole1999,Tikhonov2002,Skillman2003,Bernard2007,Skillman2014,Weisz2014,Chun2015,
Sibbons2015,Pucha2019}. 
The studies of the H II regions in this galaxy also indicate a low
metallicity of 12+log(O/H) $\simeq7.70$~\citep{Kings1995,Lee2003}. Thus, IC
1613 is known to be an extremely low metallicity galaxy, being more metal-poor than
SMC~\citep{Talent1980,Davidson1982,Dodorico1983,Peimbert1988,Herrero2010}.

On the other hand,  some studies on young stellar populations in IC 1613 imply
different metallicity values. The studies on $3\sim9$ O- and B-type supergiant
stars in this galaxy give  a metallicity of
[Z]$=-0.82\sim-0.69$~\citep{Bresolin2007,Garcia2014,Bouret2015}. More recently,
~\citet{Berger2018} found [Z]=$-0.69\pm0.24$ with  bimodal peaks at [Z]=$-0.50$
and [Z]=$-0.85$ for early B-type blue supergiants.  \citet{Taut2007}
investigated the spectra of 3 M-type RSGs and found that the average
metallicity is [Fe/H]=$-0.67\pm0.09$.  \citet{Britavskiy2019} investigated 6
RSGs including the RSG sample of~\citet{Taut2007}.  They adopted  [Fe/H]=$-0.7$
for IC 1613 and found higher effective temperatures of RSGs than those of the
SMC.  

In this study, we investigate 14 RSG stars in IC 1613 using low-resolution $J$ band
spectra obtained with the MMT and Magellan infrared spectrograph (MMIRS) on the 6.5 m
MMT telescope. We aim to estimate the effective temperatures
and metallicities of these RSGs, and calibrate the mixing length in stellar
evolution models with our RSG sample. In Section 2 we describe the RSG target
selection, observation and data reduction. In Section 3 we discuss the methods
for determining the effective temperatures and metallicity.   In Section 4 we
compare the effective temperatures of our RSG sample with stellar evolutionary
models and discuss its implications for the metallicity dependence of the
mixing length. In Section 5 we conclude this work. 

\section{Target selection, observation and data reduction} 
RSG candidates are selected from the star catalog of IC 1613 classified
by~\citet{Chun2015}. These authors investigated the stellar populations in IC
1613 using optical ($gi$) and near-infrared ($JHK_s$) photometry, and separated
the stellar populations brighter than the tip of the red-giant branch (TRGB)
into the Galactic foreground stars, supergiants, and asymptotic giant branch
(AGB) stars in IC 1613~\citep[see the text and Figure 4 in][for
details]{Chun2015}. In order to select suitable targets for MMIRS observation,
we tested the slit mask configuration 
several times to maximize the number of observable RSGs considering the sky positions of the
RSGs across the IC 1613 and distributions in the $(J-K_s, K_s)$ color-magnitude
diagram (CMD). We finally select 72 targets among the 518 RSG candidates.

\begin{figure*}
\plottwo{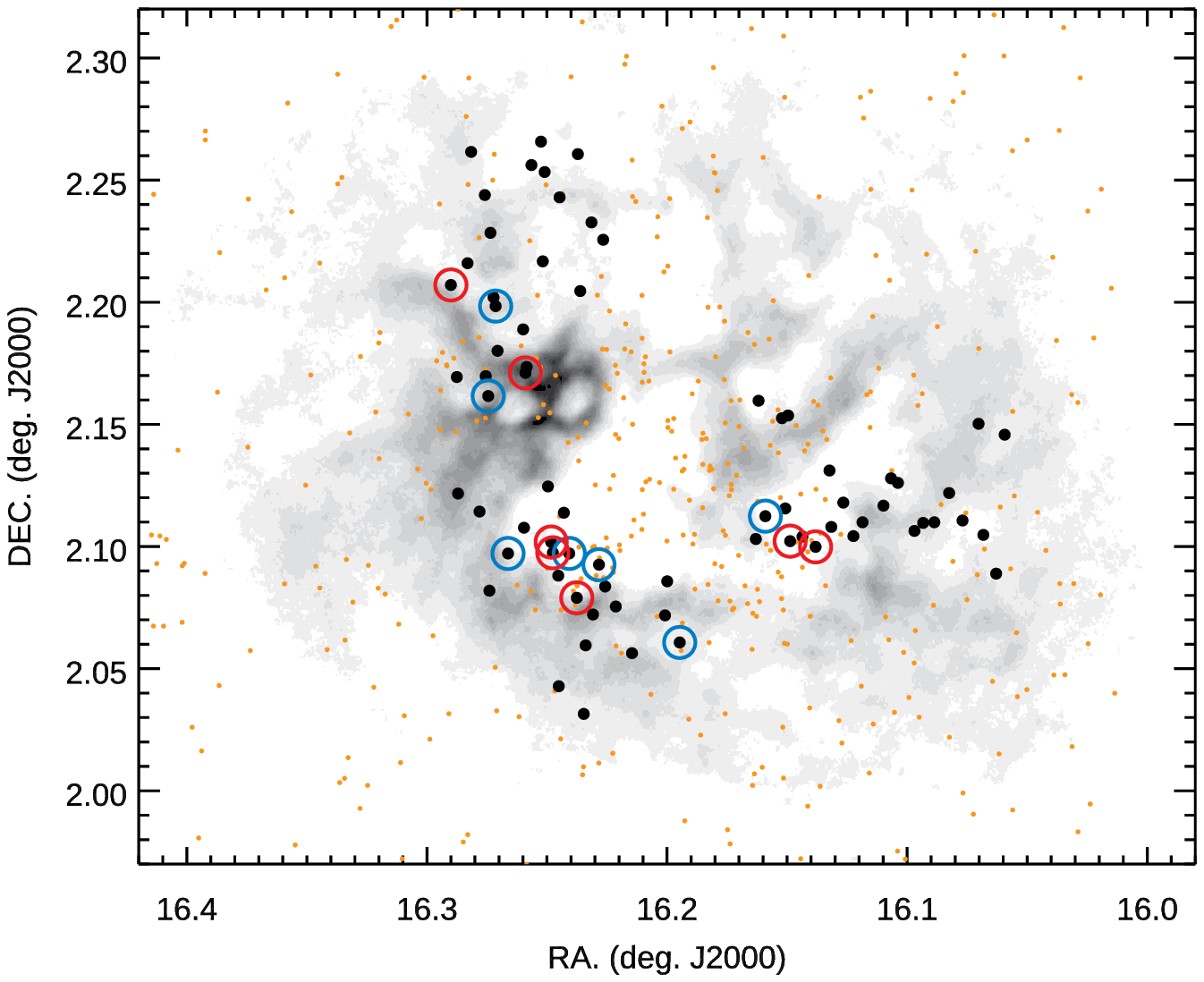}{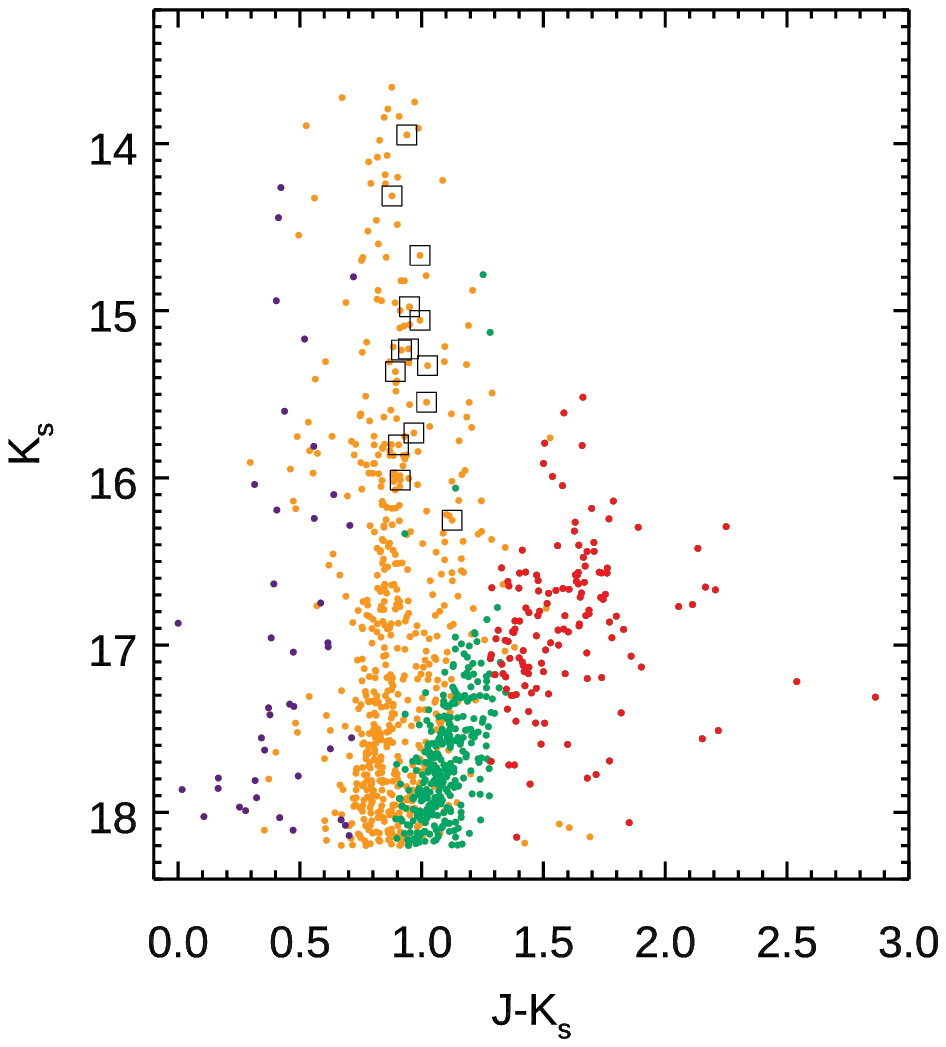}
\caption{Left panel: The sky positions of the 518 RSG candidates (orange dots) classified by
~\citet{Chun2015} and the 72 RSG candidates (black filled dots) observed in this study over
the VLA H I map of~\citet{Lozin2001}. The gray shading indicates H I column density: darker shading means higher density. 
The 14 RSGs of which the metallicity and effective temperature are inferred in this work are
indicated by open circles, and the blue and red color of circles denote the metal poor ($\mathrm{[Fe/H]}<-0.65$) and rich  ($\mathrm{[Fe/H]}>-0.65$) groups, respectively.
Right panel: $(J-K_s,K_s)$ CMD of stars brighter than the tip of red-giant branch in IC 1613 (i.e., $ K_s < 18.2$~mag). 
The green and red dots denote
M-giant and C-rich stars, respectively. The orange dots are the RSG candidates in IC 1613.
The purple dots are foreground galactic stars. The open squares are the 14 RSG targets observed with the MMIRS 
for which the metallicity and effective temperatures are inferred in this work. The photometry data and stellar classification are taken from~\citet{Chun2015}. }\label{mapcmd}
\end{figure*}

In the left panel of Figure~\ref{mapcmd}, the selected 72 RSGs candidates are
displayed on the VLA H I map of IC 1613~\citep{Lozin2001}. The RSGs in the
central region of the galaxy having the H I cavity were not chosen because of
the difficulty of the slit mask configuration due to the crowdedness. In the right
panel of Figure~\ref{mapcmd}, the $(J-K_s, K_s)$ CMD for several stellar
populations in IC 1613 classified by~\citet{Chun2015} is presented. The
preselected RSGs have a wide magnitude range, from 13.5 mag to 18.0 mag in the $K_s$
band.

The near-infrared spectra for the 72 RSG targets have been obtained using
the MMIRS instrument on the MMT telescope in the MOS mode. MMIRS has a HAWAII-2
$2048\times2048$ HgCdTe detector with a pixel scale of $0.2''$, which covers a
$4'\times7'$ field of view in the MOS mode.  The combination of $J$ grism and
$zJ$ filter, which provides $J$ band spectra spanning $0.94-1.51$ $\mu$m, was
used in the observation.  Three slit masks with a slit width of 0.5 arcsec and
a slit length of 7 arcsec to achieve a low-resolution of $R\sim2000$ were
created, following the MMT mask design procedure.  The spectra data were
acquired during the nights of September 1, 2 and 7 in 2017 by Queue mode.  Our targets
were observed using four dithering patterns with $\mathrm{+1.8}$,
$\mathrm{-1.4}$, $\mathrm{+1.4}$, and $\mathrm{-1.8}$ arcsec offsets along the
slit, and an individual exposure time of 300 seconds.  We took three
observation runs for each mask configuration to obtain enough signals.  After
the target observation, the telluric standard stars (A0V) for each mask were
observed at a similar airmass to our target in order to correct for telluric
absorption.  The log of the observation is summarized in Table~\ref{logs}.

\begin{deluxetable}{ccccc}
\tablecaption{Observation log for RSG candidates in IC 1613\label{logs}}
\tablehead{
\colhead{Mask} & \colhead{RA.} & \colhead{DEC.} & \colhead{PA.} &  \colhead{Exp. time} \\
  & (hh:mm:ss) & (dd:mm:ss) & (deg) & ($N_\mathrm{run}\times N_\mathrm{dith.} \times$ sec.)
}
\startdata
Mask1 & 01:04:58.56 & 02:05:12.43 & 42 & $3\times4\times$300 \\
Mask2 & 01:05:02.17 & 02:12:41.71 & -7 & $3\times4\times$300 \\
Mask3 & 01:04:27.12 & 02:07:19.20 & -96 & $3\times4\times$300 \\
\enddata
\end{deluxetable}

Data reduction was processed by using the MMIRS data reduction
pipeline~\citep{Chil2015} written in the IDL
language\footnote{~\url{https://bitbucket.org/chil_sai/mmirs-pipeline}}.  This
pipeline automatically performs flat fielding, wavelength calibration, sky
subtraction, and telluric correction using observed telluric standard stars for
the MMIRS spectroscopic data. The final signal-to-noise ratios (S/N) per resolution element of the
reduced spectra vary from 20 to 120, depending on the magnitudes of RSG
candidates and the weather condition. We rejected the spectra with a S/N less
than 40, and 33 RSG candidates with
an average S/N$\simeq80$ were preselected.

The radial velocities (RVs) for the 33 RSG candidates were measured by applying
a cross-correlation technique to the observed spectra in the wavelength range
from 1.15 $\mu$m to 1.25 $\mu$m, where several atomic absorption lines are
present and the molecular lines are relatively rare.  We use three template
spectra, Betelgeuse and Arcturus from the NASA Infrared Telescope Facility
spectral library for cool stars~\citep{Rayner2009}, and a synthetic spectrum.
The synthetic spectrum was generated by the local thermodynamic equilibrium
(LTE) line analysis and synthetic spectrum code MOOG~\citep{Sneden1973} with a
MARCS atmospheric model~\citep{Gusta2008} of $T_\mathrm{eff}=3800$~K,
$\mathrm{log}$ $\mathrm{g}=-0.5$ and [Fe/H]=$-0.5$.  We estimated RVs from the
cross-correlation functions (CCF) between the observed and three template
spectra, and the final RVs were averaged.  From the resulting RVs, we find that
14 RSGs having an average RV of $-160~\rm km~s^{-1}$ are hosted in IC 1613.
We note that it is difficult to precisely measure the RVs due to the low resolution and S/Ns of
our spectra.  The resulting RVs of $-160~\rm km~s^{-1}$ do not correspond to the systemic IC 1613 velocity of $-234~\rm km~s^{-1}$
measured from the H I 21 cm line spectra of~\citet{Lu1993}, but are in
agreement with the value of $-185~\rm km~s^{-1}$ for RSGs in IC 1613
by~\citet{Brita2014}.

\section{Determination of stellar parameters} 
\begin{deluxetable*}{cccccccccccc}
\tabletypesize{\footnotesize}
\tablecaption{Sky position, photometric magnitudes, and determined physical parameters of 14 RSGs in IC 1613\label{table2}}
\tablehead{
\colhead{Star} & \colhead{RA.} & \colhead{DEC.} & \colhead{$J$\tablenotemark{d}} & \colhead{$H$\tablenotemark{d}} & \colhead{$K_s$\tablenotemark{d}} & \colhead{SN} & \colhead{$T_\mathrm{eff}$} & \colhead{log~$g$} & \colhead{[Fe/H]} & \colhead{$v_{t}$} & \colhead{log L/L$_{\sun}$} \\
~ & \colhead{(hh:mm:ss)} & \colhead{(dd:mm:ss)} & ~ & ~ & ~ & ~ & \colhead{(K)} & ~ & ~ & \colhead{(km~s$^{-1}$)} & ~
}
\startdata
star1  & 1:05:03.8905 & +02:05:49.704 &  15.9264  & 15.2343  & 14.9763   &  120 &                  4310$\pm$85 (4480)              &  0.9$\pm$0.2 (0.2)   &  -0.74$\pm$0.21 (-0.86)  &  4.0$\pm$0.4 & 4.629 \\
star2  & 1:04:59.4095 & +02:05:50.784 &  15.6624  & 14.9633  & 14.6693   &  120 &                  4320$\pm$118 (4360)            &  0.8$\pm$0.3 (0.9)  &  -0.41$\pm$0.29  (-0.61)  &  3.9$\pm$0.6 & 4.729 \\
star3  & 1:04:59.5798 & +02:06:06.588 &  16.1524  & 15.4603  & 15.2353   &  100 &                  4270$\pm$85  (4390)             &  0.9$\pm$0.1 (0.9)  &  -0.52$\pm$0.17  (-1.00)  &  3.9$\pm$0.4 & 4.544 \\
star4\tablenotemark{a}  & 1:04:57.0190 & +02:04:44.400 &  16.1744  & 15.5273  & 15.2283   &  110 & 4300$\pm$170 (4420) &  0.9$\pm$0.4 (0.8)  &  -0.38$\pm$0.24  (-0.50) &  3.1$\pm$0.4 & 4.531 \\
star5  & 1:04:54.7993 & +02:05:33.216 &  16.6994  & 15.9803  & 15.7313   &  100 &                  4350$\pm$62 (4470)              &  0.6$\pm$0.7 (0.7)  &  -0.70$\pm$0.22  (-0.86) &  3.7$\pm$0.4 & 4.317 \\
star6  & 1:04:46.7303 & +02:03:38.592 &  16.2564  & 15.5903  & 15.3643   &  100 &                  4190$\pm$76 (4060)              &  0.6$\pm$0.4 (1.0)  &  -0.79$\pm$0.22  (-0.62) &  4.3$\pm$0.4 & 4.507 \\
star7  & 1:04:57.7702 & +02:05:49.812 &  16.9254  & 16.2253  & 16.0133   &    90 &                  4150$\pm$228 (4420)            &  0.7$\pm$0.6 (1.0)  &  -1.36$\pm$0.24  (-0.89) &  3.1$\pm$0.6 & 4.236 \\
star8  & 1:05:05.8704 & +02:09:41.796 &  16.0504  & 15.3303  & 15.0573   &  40 &                  4380$\pm$89 (4860)              & -0.3$\pm$0.3 (-0.2) &  -0.10$\pm$0.25  (-0.46) &  3.9$\pm$0.6 & 4.574 \\
star9  & 1:05:02.1506 & +02:10:15.888 &  16.5674  & 15.8533  & 15.5473   &  40 &                  4050$\pm$316 (4350)            & -0.4$\pm$0.5 (0.6)  &  -0.36$\pm$0.32  (-0.78) &  2.2$\pm$2.3 & 4.364 \\
star10 & 1:05:09.6002 & +02:12:25.488 &  14.8874  & 14.1993  & 13.9483   &  50 &                  4370$\pm$85 (4370)             & -0.4$\pm$0.4 (-0.5) &  -0.58$\pm$0.23   (-0.85) &  3.0$\pm$0.4 & 5.048 \\
star11\tablenotemark{b} & 1:05:05.1288 & +02:11:54.312 &  15.1914  & 14.5893  & 14.3133   &  50 & 4360$\pm$88 (4570)  & -0.3$\pm$0.4 (0.8)  &  -0.71$\pm$0.25 (-0.69) &  3.9$\pm$0.4 & 4.937 \\
star12\tablenotemark{c} & 1:04:38.1697 & +02:06:44.712 &  16.3524  & 15.6103  & 15.3283   &  50 & 4360$\pm$89 (4010)  & -0.4$\pm$0.1 (0.7)  &  -1.02$\pm$0.24  (-0.75) &  5.1$\pm$0.5 & 4.449 \\
star13 & 1:04:33.1490 & +02:05:59.496 &  16.7074  & 16.0013  & 15.8023   &  50 &                  4360$\pm$105 (4140)             &  0.9$\pm$0.2 (0.6)  &  -0.59$\pm$0.26  (-0.72) &  4.1$\pm$0.4 & 4.324 \\
star14 & 1:04:35.6905 & +02:06:07.884 &  17.3784  & 16.6543  & 16.2533   &  40 &                  4120$\pm$117 (4400)             &  0.9$\pm$0.1 (0.9)  &  -0.50$\pm$0.25  (-0.69) &  1.2$\pm$0.2 & 4.029 \\
\enddata
\tablenotetext{}{The stellar parameters obtained from lower mass models are indicated in parenthesis.}
\tablenotetext{a}{IC 1613-3 of~\citet{Britavskiy2019}}
\tablenotetext{b}{V43 of~\citet{Taut2007}}
\tablenotetext{c}{IC 1613-1 of~\citet{Britavskiy2019} }
\tablenotetext{d}{Near-infrared magnitudes come from~\citet{Chun2015}.}
\end{deluxetable*}

Inferring the effective temperature ($T_\mathrm{eff}$) is needed to understand
the physical properties of RSGs and to confront the predictions of stellar
evolution models, but it has been a challenging issue.
\citet{Levesque2005,Levesque2006} present comprehensive studies on the
effective temperature of RSGs.  They derive  $T_\mathrm{eff}$ of RSGs in the
Milky Way and the Magellanic Clouds by fitting the synthetic spectra generated
with the MARCS atmosphere models to the TiO absorption band of the optical
spectra. On the other hand, ~\citet{Davies2013}  estimate $T_\mathrm{eff}$ of
RSGs in the Magellanic Clouds using the spectral energy distributions (SEDs),
finding that  $T_\mathrm{eff}$ inferred from the TiO band ($T_\mathrm{eff, TiO}$) is systematically
lower than that from the SEDs ($T_\mathrm{eff, SED}$).  \citet{Davies2013} argue that the TiO band
property could not reflect the exact $T_\mathrm{eff}$ of the RSGs because the
relatively upper layer of the atmosphere where the TiO band forms might be
significantly affected by three-dimensional effects, such as granulation.
Regarding the origin of the difference between $T_\mathrm{eff, TiO}$ and $T_\mathrm{eff, SED}$,
~\citet{Davies2021} recently suggest that the TiO absorption becomes stronger with presence of a
strong wind, shifting the spectral type of a RSG to a later type for a given $T_\mathrm{eff}$.
Alternatively atomic lines produced at a layer below the TiO band forming
region would be less affected by the three-dimensional
effects~\citep{Davies2013,Tabernero2018}.  
Using atomic lines to infer
$T_\mathrm{eff}$ and other stellar parameters of RSGs is investigated by
several studies:  CaT
features~\citep{Dorda2016a,Dorda2016b,Tabernero2018,Dicenzo2019}, $J$-band
technique~\citep{Gazak2014b,Davies2015}, and line-depth ratios of iron
lines~\citep{Taniguchi2018,Taniguchi2020}.  However, there are still some
systematic offsets between the derived $T_\mathrm{eff}s$ from different methods. 

In the present study, we apply the $J$-band technique to determine the stellar
parameters of our RSG sample in IC 1613.  We used the spectra ranging from 1.16
$\mu$m to 1.23 $\mu$m, where several atomic absorption lines (e.g., Fe, Mg, Si,
and Ti) are present with little contamination by molecular lines, to apply the
$J$-band technique discussed by~\citet{Gazak2014b} and~\citet{Davies2015}.  To
make a spectral fitting to the observed spectra, a synthetic spectra grid is
generated using the online spectrum
tool\footnote{~\url{http://nlte.mpia.de/index.php}}~\citep{NLTE} hosted by the
Max Planck Institute for Astronomy (MPIA), for which the RSG MARCS atmospheric
models and the solar-scaled chemical abundance ratio from~\citet{Grevesse2007}
are used.  The NLTE effects for several elements (e.g., Fe, Si, Ti, and Mg) are
corrected~\citep{Bergemann2012,Bergemann2013,Bergemann2015}.  The grid covers a
$T_\mathrm{eff}$ range of 3400 K $\sim$ 4400 K in increments of 100 K.  The
surface gravity of log $g$ ranges from -0.5 to +1.0 in increments of 0.5 (in
cgs units), and the metallicity of [Fe/H] from -1.5 dex to +1.0 dex in
increments of 0.25 dex.  The microturbulence ($\xi$) range is from 1.0
km~s$^{-1}$ to 6.0 km~s$^{-1}$ in 1.0 km~s$^{-1}$ increments.

The generated synthetic spectra are compared with the observed spectra, and
stellar parameters and metallicity are determined by using $\chi^2$
minimization. As a first step, we fit the continuum level of both the observed
and model spectra. We set wavelength points where any given synthetic spectrum
within a 2--5 $\mathrm{\AA}$ wavelength window has a maximum flux, and we
calculate the ratios of the model to the observed spectra flux at these
wavelength points assuming the continuum level. The continuum
correction function is then constructed by fitting the ratios with a low-order
polynomial function. The outlier points with the residual of the fit larger than
3$\sigma$ are rejected. The final continuum correction function is applied to
the synthetic spectra to fit the continuum level of the observation.  At the
same time, the synthetic spectra are also convolved to match the observed line
profiles and the spectral resolution.

\begin{figure}
\includegraphics[width=\columnwidth]{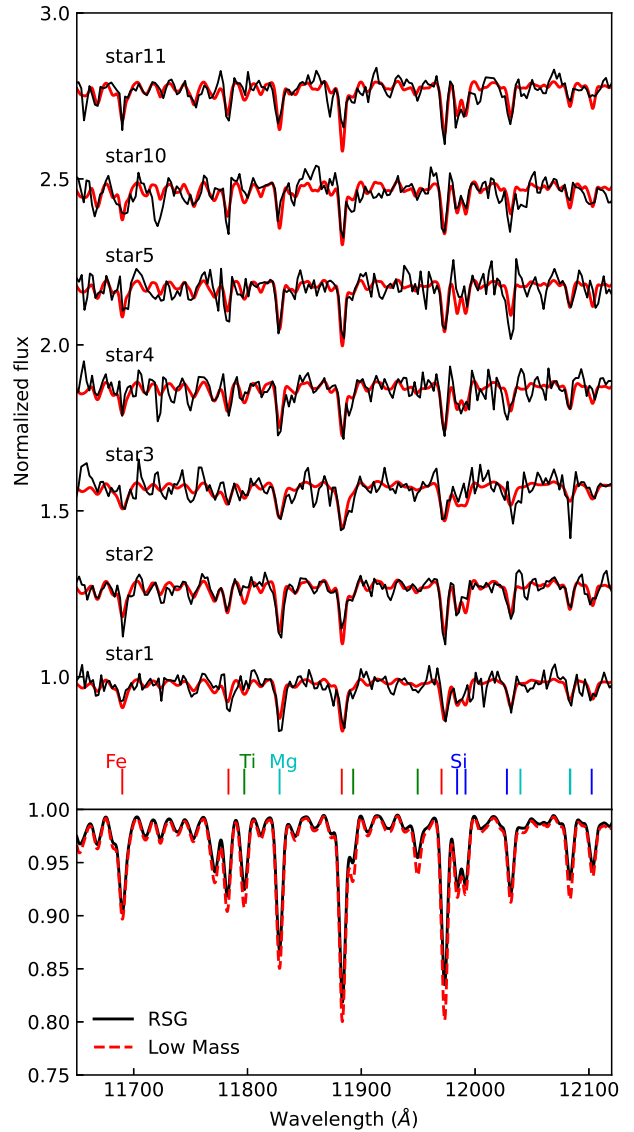}
\caption{Top: Sample spectra of the observed RSG targets (black lines) with the best fit
synthetic spectra (red lines). The prominent atomic lines are indicated by vertical lines.
Bottom: The comparison between RSG model of 15$M_\sun$ (black) and low-mass model of 1$M_\sun$(red) spectra at low metallicity ([Fe/H]=$-0.7)$.
}\label{spec}
\end{figure}

The $\chi^2$ values for all synthetic spectra are calculated, and
the best model with the lowest $\chi^2$ value was selected.  As investigated
by~\citet{Gazak2014b}, we keep two parameters among the four stellar parameters ([Fe/H], $T_\mathrm{eff}$, log $g$ and  $\xi$)
of the selected best model and investigate the $\chi^2$ distributions by varying the remaining two free parameters (e.g., [Fe/H] - $T_\mathrm{eff}$, [Fe/H] - log $g$, etc.).
In this process, we can make six $\chi^2$ distribution planes with a combination of two different free parameters.
For each of the six planes, we interpolate the $\chi^2$ grid of the two free
parameters onto a denser grid plane and take two parameters at the minimum $\chi^2$ value as best
fit values. Three best fit values for each stellar parameter were obtained from six $\chi^2$ planes.
The final best fit parameters are averaged by these three values for each parameter. Figure~\ref{spec} shows
examples of the observed RSG spectra with the best fit synthetic spectra.

We estimate the uncertainties in deriving stellar parameters through Monte
Carlo simulations.  The synthetic spectra of the final best fit parameters are
interpolated in the online spectrum tool with the correction of the NLTE
effect.  For each synthetic spectrum, we generate 200 noisy spectra by adding
a random Gaussian noise to simulate the S/Ns of the observed spectrum.  The best
fit parameters of the individual noisy spectra are calculated through our
analysis, and the distributions of the minimum $\chi^2$ for each parameter are
investigated.  The iso-contour levels, which encompass $1\sigma$ around a
minimum $\chi^2$, are calculated, and then we take the minimum and maximum
parameter values within the iso-contours to access $1\sigma$ uncertainty in
stellar parameters. We summarize the fundamental physical parameters for the
14 RSG targets in IC 1613 in Table~\ref{table2}.

Although  previous studies for RSGs in IC 1613 are limited, we compare
our stellar parameters with the results of previous studies.  We find that
three RSGs (star4, star11, and star12) in our RSG targets are commonly
detected in previous RSG studies: IC 1613-1 and IC 1613-3
in~\citet{Britavskiy2019} and V43 in~\citet{Taut2007}. The cross-identified
RSGs in previous studies are indicated in the note in Table~\ref{table2}. We
find that the average difference in the effective temperature ($\bigtriangleup
T_\mathrm{eff}$) is about 350 K, which is so large that our results seem to be
inconsistent with previous results.  However, we note
that~\citet{Britavskiy2019} derived effective temperatures of RSGs using model
SEDs with a fixed metallicity of [Fe/H]=$-1.0$ which is much lower than the
metallicity measured in this study and~\citet{Taut2007}, and also used
narrow optical spectra with low S/N ratio which result in the significant uncertainties
of the SED fitting.

\subsection{Sensitivity test for RSG models, Signal-to-noise ratio, and spectral resolution}
\begin{figure*}
\gridline{\fig{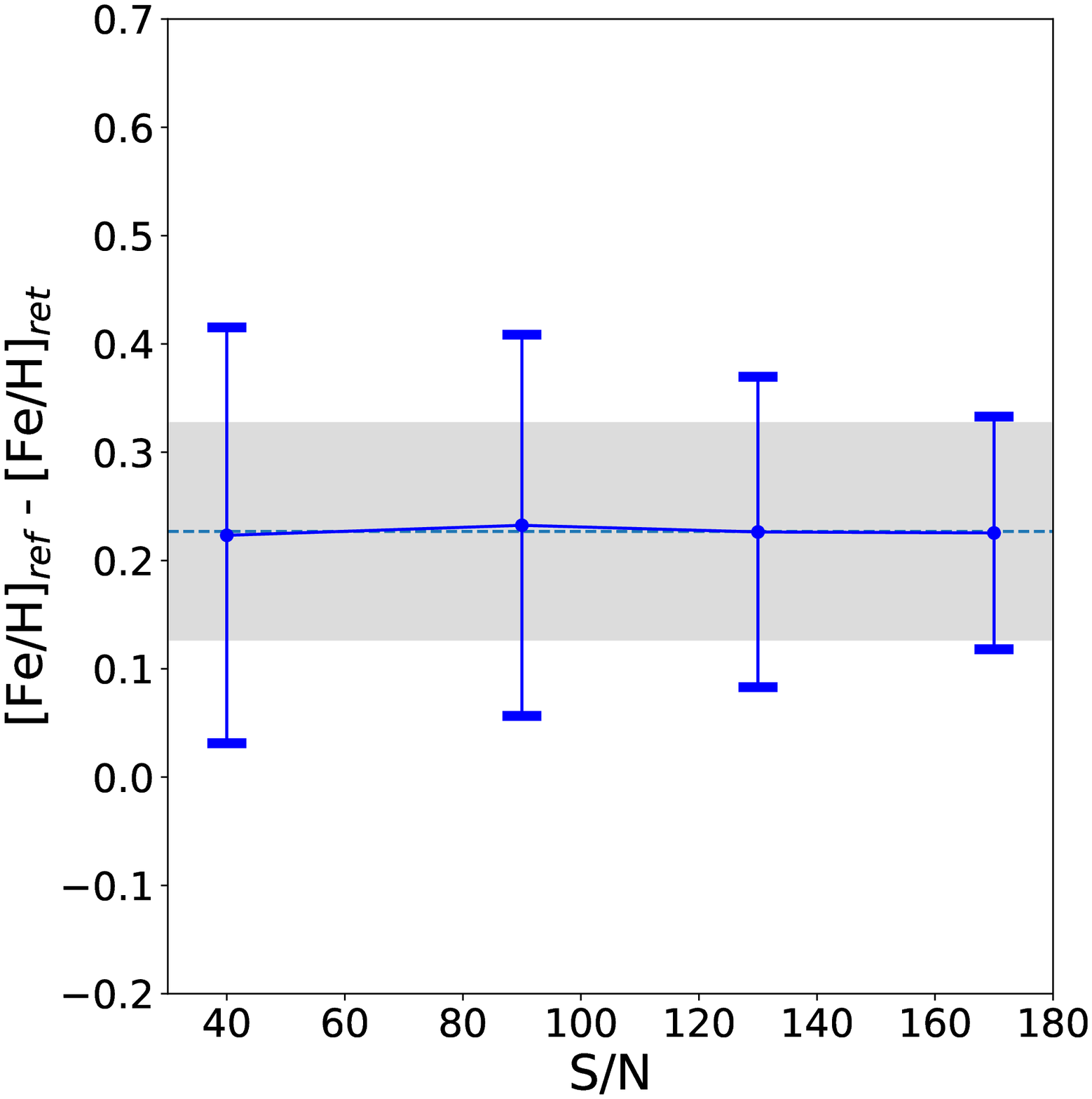}{0.4\textwidth}{(a)}
            \fig{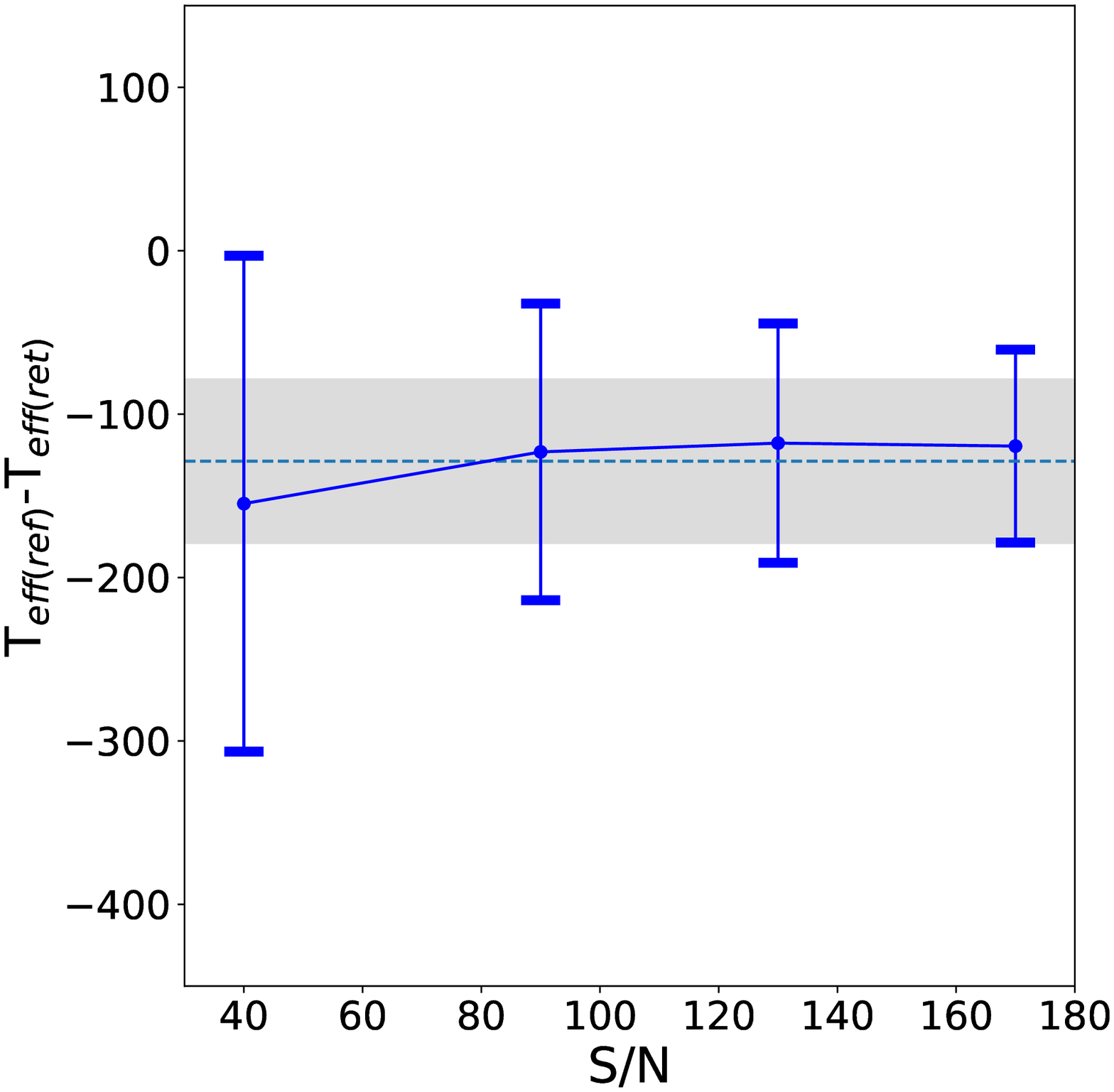}{0.4\textwidth}{(b)}}
\gridline{\fig{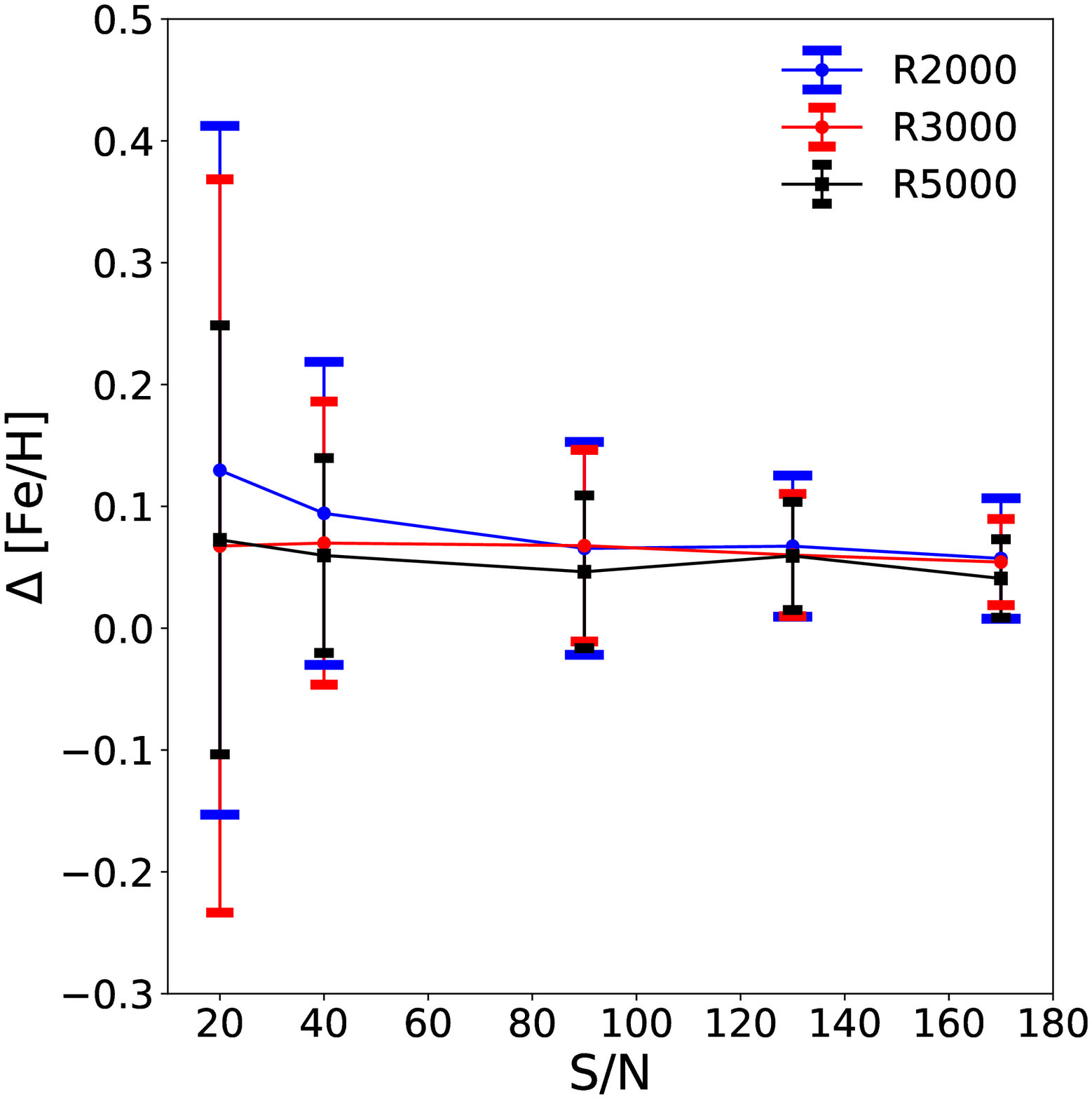}{0.4\textwidth}{(c)}
            \fig{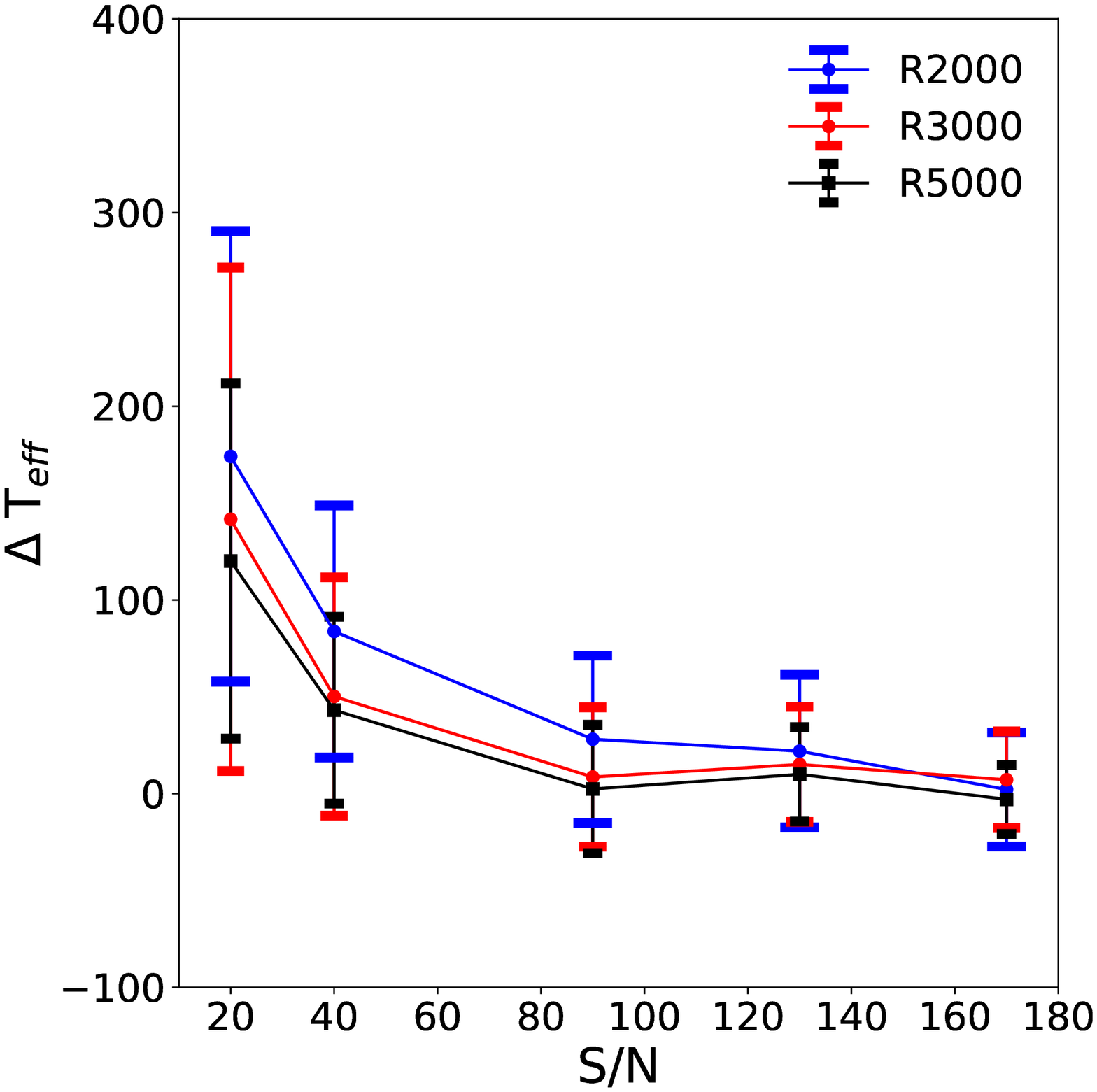}{0.4\textwidth}{(d)}}
\caption{Top panels: The systematic offsets in metallicity (a) and temperature (b) obtained from low-mass spectra to RSG reference spectra 
at low metallicity ([Fe/H]=$-0.7$). The dotted lines are average offsets in metallicity and temperature and shaded area is $1\sigma$ error of average values.
Bottom panels: The differences of metallicity (c) and effective temperature (d) between input and retrieved stellar parameters as a function of S/N per resolution elements and spectral resolution.
The different resolutions are indicated by different colors: blue for R=2000, red for R=3000, and black for R=5000.
 }\label{SNR}
\end{figure*}
We found that the obtained
effective temperatures of RSGs seem to be close to the edge ($\sim 4400$ K) of the model grid.
The narrow grid coverage might introduce a bias into the results.
A grid with a broader temperature range that fully encompasses the observed values is needed to obtain a more reliable result.
Unfortunately, however, current available RSG models are limited up to $T_\mathrm{eff}=4400$ K temperature.

Regarding this problem, ~\citet{Davies2010,Bergemann2012} noted that the model spectra between 1-5$M_\sun$ and 15$M_\sun$ for the same
stellar parameters do not show much difference and there are small differences in NLTE abundance correction, according to which 
spherical MARCS models with lower mass instead of 15$M_\sun$ RSG models could be used to derive the stellar parameters of the RSGs.
Contrary to the note of~\citet{Davies2010,Bergemann2012}, however, we found that there are considerable differences
in the intensity of absorption lines between low-mass of 1$M_\sun$ and RSG of 15$M_\sun$ spectra at low metallicity (i.e., [Fe/H]=$-0.7$)
across the $J-$band as shown in the bottom panel of Figure~\ref{spec}. 
The spectra of the low-mass star model show relatively stronger absorption lines than RSG spectra, which would lead to higher temperature and lower metallicity when we use the low-mass model spectra to fit the observed RSGs. 
Therefore, we calculate systematic offsets in temperature and metallicity in using low-mass models to derive RSG stellar parameters.
Two sets of 100 RSG reference model spectra with $T_\mathrm{eff}=4000$ K and $4300$ K at IC 1613 metallicity were made, and then
the stellar parameters of these spectra were estimated by using low-mass model spectra.
The effective temperature range of $3500 \leq T_\mathrm{eff} \leq 5000$ K and a fixed microturbulence ($\xi$) of 4.0 km s$^{-1}$ were used for model spectra.

In the top panels (a) and (b) in Figure~\ref{SNR}, systematic offsets in metallicity and temperature 
obtained from low-mass models to the RSG reference spectra are plotted as a function of S/N ratios.
We note that there is no a significant trend in the calculated systematic offsets between $T_\mathrm{eff}=4000$ K and $4300$ K RSG reference spectra.
We find that the average temperature obtained from low-mass model spectra is systematically higher by about $130\pm50$ K than
that of reference RSG models, while the metallicity is lower by about $0.2\pm0.1$ dex.
This result is consistent with the predicted trend from using low-mass model, and 
corrections using these offset values should be applied to the stellar parameters obtained from the low-mass model spectra.

With the correction values, we use low-mass models and derive again the stellar parameters of the observed RSGs using the same fitting process.
The stellar parameters after systematic corrections (i.e., $\triangle T=130$ K and $\triangle$ [Fe/H]$=0.2$) are indicated 
in the parenthesis in Table~\ref{table2}. 
The final corrected average temperature 
and metallicity (i.e., $\langle T_\mathrm{eff} \rangle = 4380\pm200$ K, $\langle \mathrm{[Fe/H]} \rangle$ = $-0.73\pm0.15$) are comparable to those ($\langle T_\mathrm{eff} \rangle = 4280\pm110$ K, $\langle \mathrm{[Fe/H]} \rangle$ = $-0.68\pm0.28$) obtained from RSG model spectra.
We note that the average temperature and metallicity without corrections are higher by about 180 K and lower by 0.18 dex than
those obtained from RSG model spectra, respectively. 
The differences in temperature and metallicity are almost consistent with the predicted systematic offsets. 
Therefore, we adopt and use the stellar parameters obtained from RSG models in the following analysis and figures.

~\citet{Gazak2014b} has tested the dependence of the inferred stellar parameters obtained by the J-band technique 
on the spectral resolution and S/N for the RSG spectra of solar metallicity, and shown that
low resolution and/or low S/N can introduce systematic errors into the measured temperature and metallicity. 
~\citet{Gazak2014b} suggested that the stellar parameters with reliable accuracy could be obtained
from the RSG spectra with S/N $>$ 100 at the minimum spectral resolution of R=2000.
Thus we also test the sensitivity of our stellar parameters as functions of spectral resolution and S/N.
We made sets of 60 metal-poor ([Fe/H]= $-0.7$) RSG model spectra at given resolutions (R=2000, 3000, and 5000) and
S/N per resolution elements (as 20, 40, 90, 130, and 170), respectively, and looked into how the retrieved stellar parameters compare to the input parameters.
The average and standard deviation of the difference between the input and retrieved parameters were calculated.

The bottom panels (c) and (d) of Figure~\ref{SNR} show the differences of metallicity and effective temperature between input and retrieved
parameters as functions of SN ratios and resolutions.
The high-resolution spectra with a higher SN ratio provide more accurate effective 
temperatures with a small standard deviation. 
However, the temperature difference between the results with SN=40 and 100 at R=2000 is about 80 K, while the 
variation between the results with R=2000 and 5000 at SN=40 is about 40 K.
In addition, the metallicity does not seem to be significantly changed by the resolution and SN ratio.
For the spectral resolution (R=2000) and the SN ratio of our spectra, the maximum
difference from the result of a higher resolution/SN ratio would be about 0.2 dex in metallicity and 150 K in temperature.
Therefore we conclude that the resolution and SN ratio of our spectra can provide reasonable stellar parameters.

\subsection{Metallicity and temperature distributions}
\begin{figure}
\includegraphics[width=\columnwidth]{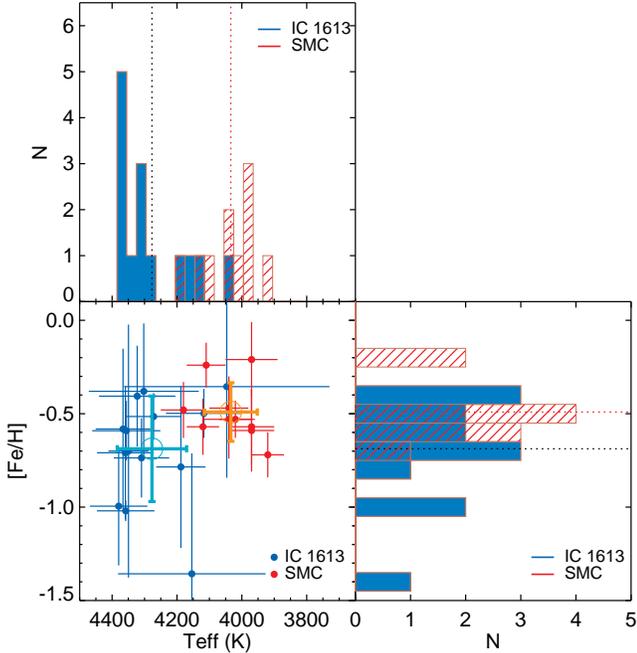}
\caption{The effective temperatures of RSGs in IC 1613 as a function of metallicity.
The RSG data in SMC given by ~\citet{Davies2015} are also plotted with red dots.
The mean metallicities and temperatures for IC 1613 and SMC are marked by cyan and
orange open dots, respectively.
The temperature and metallicity distributions are presented in the upper and right
panels. The blue and red-hatched histograms are for IC 1613 and SMC, respectively.
The black and red dotted lines in the distribution panels indicate
the mean values for IC 1613 and SMC, respectively.   
}\label{tempmetal}
\end{figure}

We present the metallicity and effective temperature of our 14 RSG targets in
IC 1613 in Figure~\ref{tempmetal}, compared with those of the RSGs in SMC given
by \citet{Davies2015}. The average metallicity of the RSG sample in IC 1613 is
[Fe/H]=$-0.69\pm0.27$, which is slightly lower than the average metallicity
([Fe/H]=$-0.49\pm0.16$) of the SMC sample  of \citet{Davies2015}.  This
metallicity value is significantly higher than the metallicity of old
and intermediate-age stellar populations in IC 1613
~\citep[$\mathrm{[Fe/H]}\sim-1.9$ to
$-1.19$][]{Freedman1988b,Cole1999,Tik2002,Skillman2003,Kirby2013} but
consistent with the young stellar population metallicity of
[Fe/H]=$-0.69\pm0.09$~\citep{Taut2007}.  \citet{Berger2018} also reports
[Z]=$-0.69\pm0.24$  for blue supergiant stars in IC 1613, which are RSG
precursors.  Since IC 1613 has been continuously star forming at a constant
rate throughout its lifetime~\citep{Cole1999,Bernard2007,Skillman2014},
the metal-richness of young stellar populations compared to old
and intermediate-age stellar populations is expected, as also discussed in  
~\citet{Taut2007}.

One of the interesting results of~\citet{Berger2018} is the bimodal metallicity
distribution of blue supergiants and the spatial concentration of the metal-rich
component in the central region of the galaxy. The metallicity
distribution in our RSG sample shows a broad spread (see Figure~\ref{tempmetal}). 
However, 
local peaks are also found at around [Fe/H]$\sim-0.4$ and [Fe/H]$\sim-0.7$, which
are almost the same as those of~\citet{Berger2018}.  Therefore, we divided the
RSGs into two groups; 7 metal-rich ([Fe/H]$>-0.65$) and 7 metal-poor
([Fe/H]$<-0.65$).  In the left panel of Figure~\ref{mapcmd}, the metal-rich and
metal-poor groups are indicated by red and blue open circles. Although the RSGs
in the center of the galaxy were not observed in this study, there seems to be
no spatial dependence between the metal-rich and metal-poor groups.  
Regarding the metallicity dependence, a
non-significant~\citep{Sibbons2015} or slightly negative metallicity
gradient~\citep{Chun2015} as a function of radial distance is reported for the
intermediate-age AGB stars.  Therefore, a study on metallicity of the
RSGs in the central region or the H I cavity as well as metal-rich early A-type supergiants (ASGs) of~\citet{Berger2018} is necessary 
to confirm the bimodal metallicity distribution and spatial dependence of RSGs. 

Stellar evolutionary models predict systematically lower RSG effective
temperatures at higher metallicity for a given mixing length
\citep[see][for a recent discussion]{Chun2018}.  In the effective
temperature distribution of Figure~\ref{tempmetal}, we find that the 
average temperature of RSGs in IC 1613 is higher (about 250 K) than that of
RSGs in the SMC.
Note, however, that no clear correlation between $T_\mathrm{eff}$ and metallicity among 
our RSG sample is found. The spectroscopic data with the higher resolution are necessary
to find such a correlation among the RSGs in the galaxy.
The trend of increasing temperature toward lower metallicity is also reported in the study of~\citet{Britavskiy2019}. 
They compared the effective temperatures of a
small number of RSGs in several dwarf irregular galaxies (including IC 1613)
having different metallicity environments in the Local Group, finding a clear
trend of an increasing RSG effective temperature toward lower metallicity.  

Our results appear to be in contrast to the previous work from~\citet{Davies2013,Davies2015}, who find
roughly uniform temperature distribution of RSGs in the LMC and SMC. 
However, we have to note that they just find no systematic difference in temperature between the LMC ($4170\pm170$ K) and SMC ($4030\pm90$ K)
for a small sample (9-10 RGSs) at the level of the precision of the measurements ($\sim 190$ K). 
~\citet{Davies2018} could later find a systematic shift in spectral types and temperature 
for a large sample of cool supergiants in LMC and SMC.
~\citet{Tabernero2018} also find a significant difference ($\bigtriangleup T \sim 150$ K, see Figure 3 in their paper) in the temperature distributions between
LMC and SMC for more than 400 RSGs.
Although ~\citet{Gonzalez2021} could not find a meaningful difference in temperature between the LMC ($4140\pm148$ K) and SMC ($4130\pm103$ K) 
because of the rather large uncertainty ($\sim 180$ K) in their temperature measurement,
they find that RSGs in the WLM dwarf galaxy, which has a lower metallicity than SMC, have a lower average $T_\mathrm{eff}$ than 
RSGs in SMC by about 250 K.
Therefore, our finding that the average $T_\mathrm{eff}$ of RSGs in IC 1613 is higher than that of SMC RSGs supports the conclusion of 
~\citet{Davies2018}, ~\citet{Tabernero2018},  and ~\citet{Gonzalez2021} that 
$T_\mathrm{eff}$ of RSGs increases with decreasing metallicity.
A more accurate investigation of metallicity and temperature for a
larger sample of RSGs in IC 1613 is needed to confirm the metallicity dependence of the RSG temperature in environments more 
metal-poor than the SMC.

\section{Comparison with the stellar evolutionary tracks}
\begin{figure*}
\includegraphics[width=\textwidth]{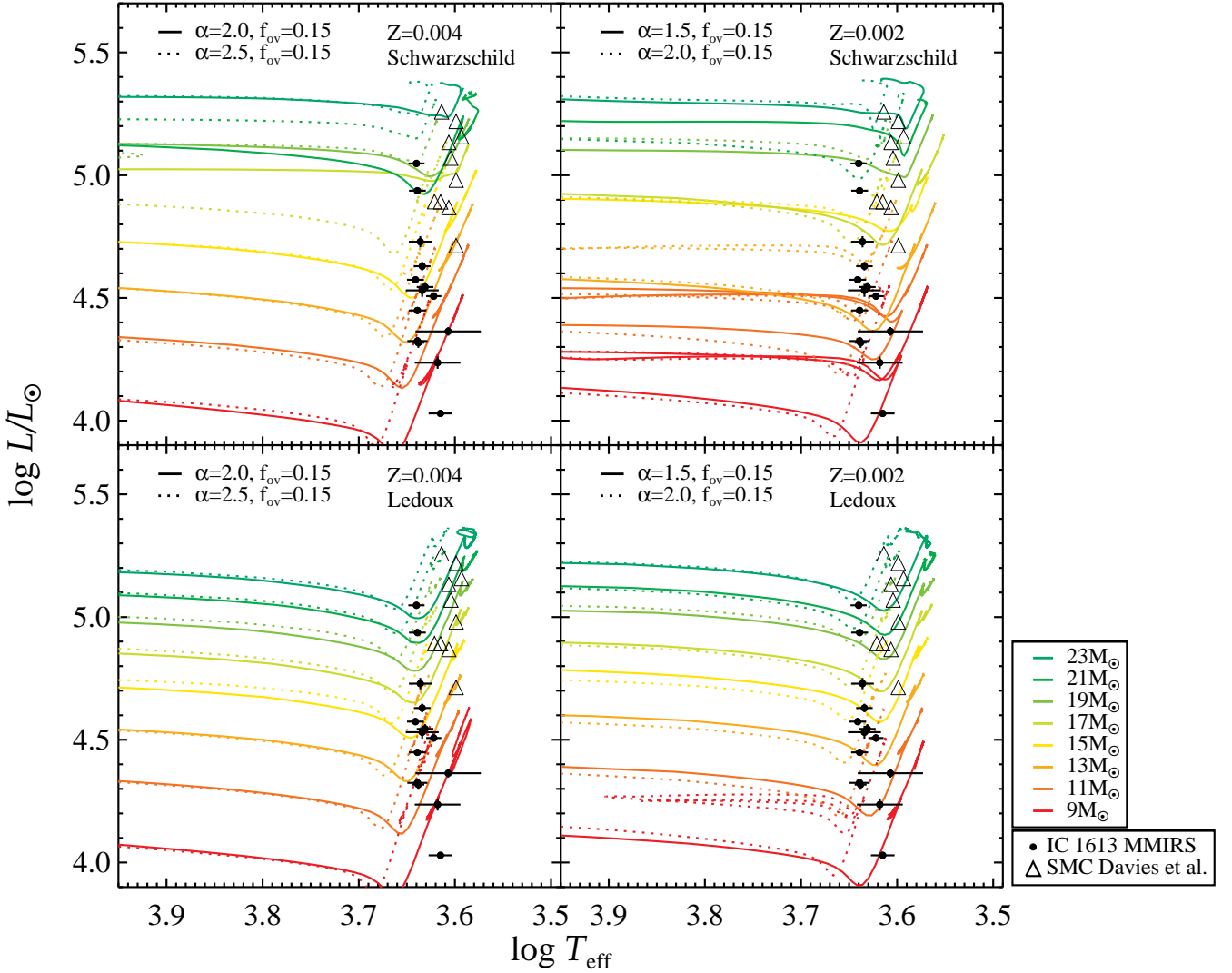}
\caption{RSGs of IC 1613 on the HR diagram compared with
the evolutionary tracks of SMC-like metallicity ($Z=0.004$, left panel) and $Z=0.002$ (right panel) with $f_\mathrm{ov}=0.15$, respectively. 
The Schwarzschild and Ledoux models are plotted in the top and bottom panels, respectively.
The tracks with mixing lengths of $\alpha_{\mathrm{MLT}}=2.0$ (solid line) and $\alpha_{\mathrm{MLT}}=2.5$ (dotted line)
are presented for $Z=0.004$, and those with $\alpha=1.5$ (solid line) and $\alpha=2.0$ (dotted line) for $Z=0.002$.
The initial mass of each track is indicated by the color of the line.
The RSGs in SMC from~\citet{Davies2015} are indicated by open triangles.
}
\label{HR}
\end{figure*}
We derive the bolometric luminosities of RSGs and compared the derived
temperatures and luminosities with the evolutionary tracks in a
Hertzsrung-Russel diagram (HRD). We use the near-infrared photometric
magnitudes of~\citet{Chun2015} in the $K$-band because the WIRCam photometry
by~\citet{Chun2015} provides more accurate magnitudes than those of the Two
Micron All Sky Survey~\citep[2MASS,][]{Cutri2003}. 
The near-infrared magnitudes of WIRCam photometry are presented in Table~\ref{table2}.
We note that the magnitudes
of all RSG candidates of IC 1613 are on or below the limit of 2MASS. The
bolometric magnitudes are calculated using the extinction values
from~\citet{Schlafly2011}, the bolometric correction relations between the
$K_s$ band and $J-K_s$ for the spectroscopically late-type long periodic
variables~\citep{Bessell1984}, and the distance modulus of
$\mu_0=24.291$~\citep{Piet2006}.  This method is relatively insensitive to
extinction as reported by~\citet{Britavskiy2019}. 
We also derived the bolometric luminosities using bolometric correction 
of~\citet{Davies2013}, and found that the average difference in both luminosities is 
about 0.005 dex.

The evolutionary tracks to compare with our RSG targets are calculated with
the MESA code~\citep{Paxton2011,Paxton2013,Paxton2015}.  We follow the same
physical assumptions described in Sect. 2 of~\citet{Chun2018}.  In short, we
considered both the Schwarzschild and Ledoux criteria for convection.  Since the
average metallicity of RSGs in IC 1613 is  comparable to or slightly lower than that 
of SMC, we calculate the evolutionary tracks of the SMC-like
($Z=0.004$) and lower ($Z=0.002$) metallicities.  For each metallicity, four
different mixing length parameters: $\alpha_{\mathrm{MLT}}=1.5, 2.0, 2.5$ and $3.0$, which are
given in units of the local pressure scale height, are considered.  An
overshooting parameter of $f_\mathrm{ov}=0.15$, which is given in units of
local pressure scale height at the upper boundary of the convective core, is
used.

In Figure~\ref{HR} we present the RSGs of IC 1613 on the HR diagram compared
with the evolutionary tracks at $Z=0.004$ and $Z=0.002$.  We also include the
RSGs in the SMC of~\citet{Davies2015} in the figure.  As discussed
in~\citet{Chun2018}, the evolutionary tracks of $Z=0.004$ with about
$\alpha_{\mathrm{MLT}}=2.0$ for both the Schwarzschild and Ledoux criteria can reproduce the
temperatures of the RSGs in the SMC. However, these tracks are systematically
cooler than the temperatures of the RSGs in IC 1613. The evolutionary tracks
with $\alpha_{\mathrm{MLT}}=2.5$ are roughly compatible to the positions of the RSGs of IC
1613.  For $Z=0.002$, the evolutionary tracks with $\alpha_{\mathrm{MLT}}=2.0$ 
give effective temperatures consistent with those of observed RSGs in IC 1613.  

The average luminosity of our RSG targets in IC 1613 is lower than that of the
SMC.  The majority of RSGs in IC 1613 are located below the 19M$_\sun$
evolutionary track (Figure~\ref{HR}).  This is simply because of the
selection effect in our work. As shown in the CMD of Figure~\ref{mapcmd}, the
majority of the observed RSGs in IC 1613 have a magnitude of $K_s>15$, and
there are many bright RSG candidates that are not observed in this study.  If
we had a larger sample of RSGs including bright candidates over the full
spatial extent of IC 1613, it would be possible to constrain the
$L_\mathrm{max}$ or the Humphreys-Davidson limit in this galaxy.

\begin{figure*}
\includegraphics[width=0.9\textwidth]{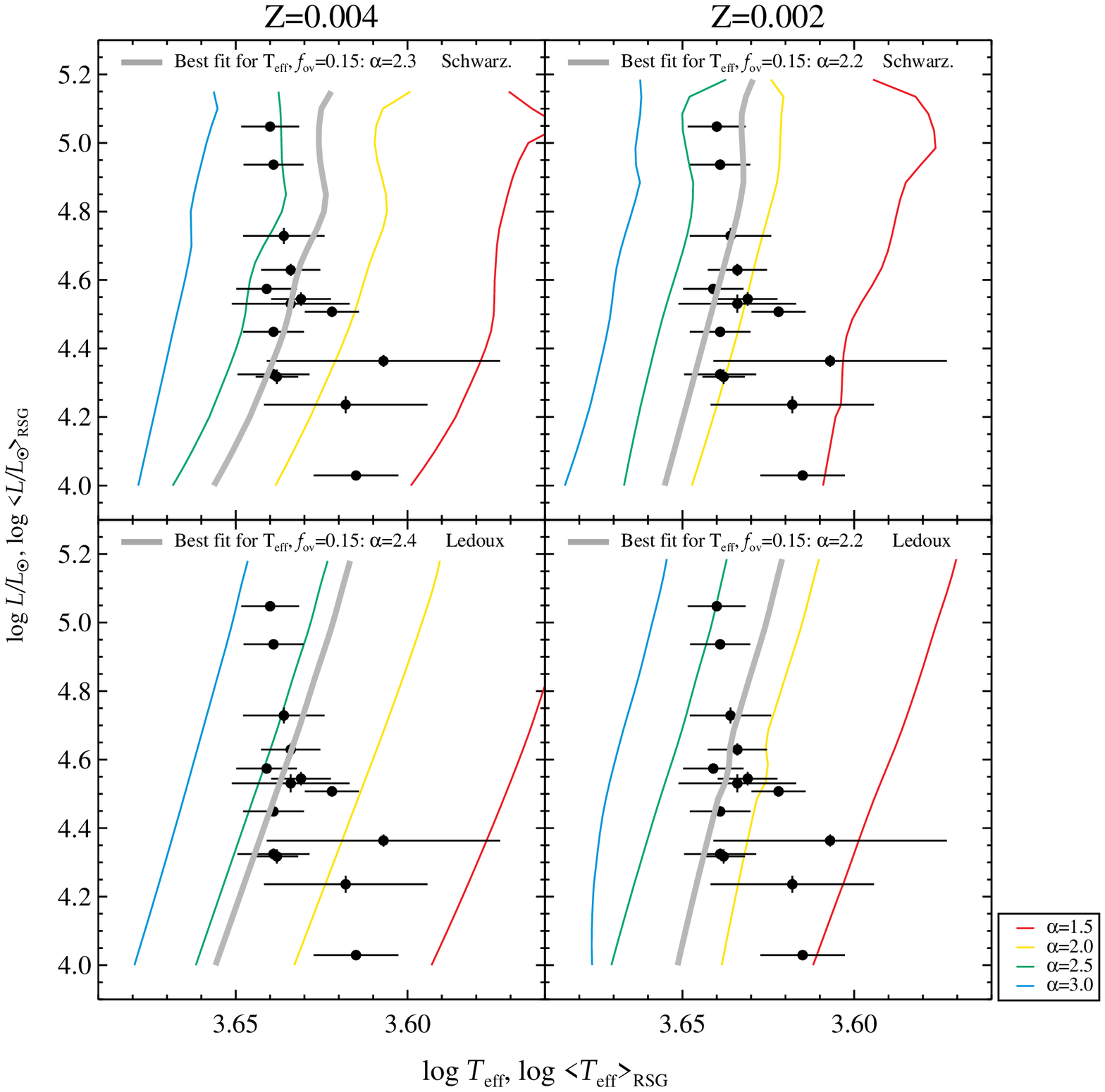}
\caption{RSGs of IC 1613 on the HR diagram compared with
the time  weighted temperatures and luminosities of the evolutionary tracks with
metallicities of Z=0.004 (left panel) and 0.002 (right panel). 
The Schwarzschild and Ledoux models are plotted in the top and bottom panels, respectively.
The time-weighted values of tracks for different mixing lengths ($\alpha_{\mathrm{MLT}}=1.5, 2.0, 2.5,$ and $3.0$)
are indicated by red, yellow, green, and cyan lines, respectively. The best fits to the derived effective
temperatures of RSGs in IC 1613 are represented by thick gray line and the mixing length
that gives the best fit is indicated by the label in the upper part of each panel. 
}
\label{aveHR}
\end{figure*}

In order to find the mixing length value that gives the best fit to the
position of RSGs in the HR diagram, the time-weighted effective temperature
($<T_\mathrm{eff}>_\mathrm{RSG}$) and luminosity ($<L>_\mathrm{RSG}$) from the
evolutionary tracks are calculated.  We interpolate
$<T_\mathrm{eff}>_\mathrm{RSG}$ and $<L>_\mathrm{RSG}$ at mixing lengths from
$\alpha=1.5$ to 3.0 in increments of 0.1, and then compare the effective
temperatures of the RSGs in IC 1613 with those of the interpolated values at a
given luminosity. The $\chi^2$ value is calculated from the deviation between
the observation and model temperatures, and the mixing length value with the
lowest $\chi^2$ is selected as the best fit value~\citep[see  
Sect. 4 of][for more details on this approach.]{Chun2018}.

In Figure~\ref{aveHR}, the effective temperatures and luminosities of the RSGs
in IC 1613 are compared with time-weighted temperatures and luminosities of the
evolutionary tracks in the HR diagram.  From our $\chi^2$ minimization
analysis, we find that the best fits of the Schwarzschild and Ledoux models
with $Z = 0.004$ to the observation are given by $\alpha_{\mathrm{MLT}}=2.3$ and $2.4$,
respectively. On the other hand, for $Z=0.002$, a lower mixing length value of
$\alpha_{\mathrm{MLT}}=2.2$ for both the Schwarzschild and Ledoux models gives the best fit to
the observed data. 

\begin{figure}
\includegraphics[width=\columnwidth]{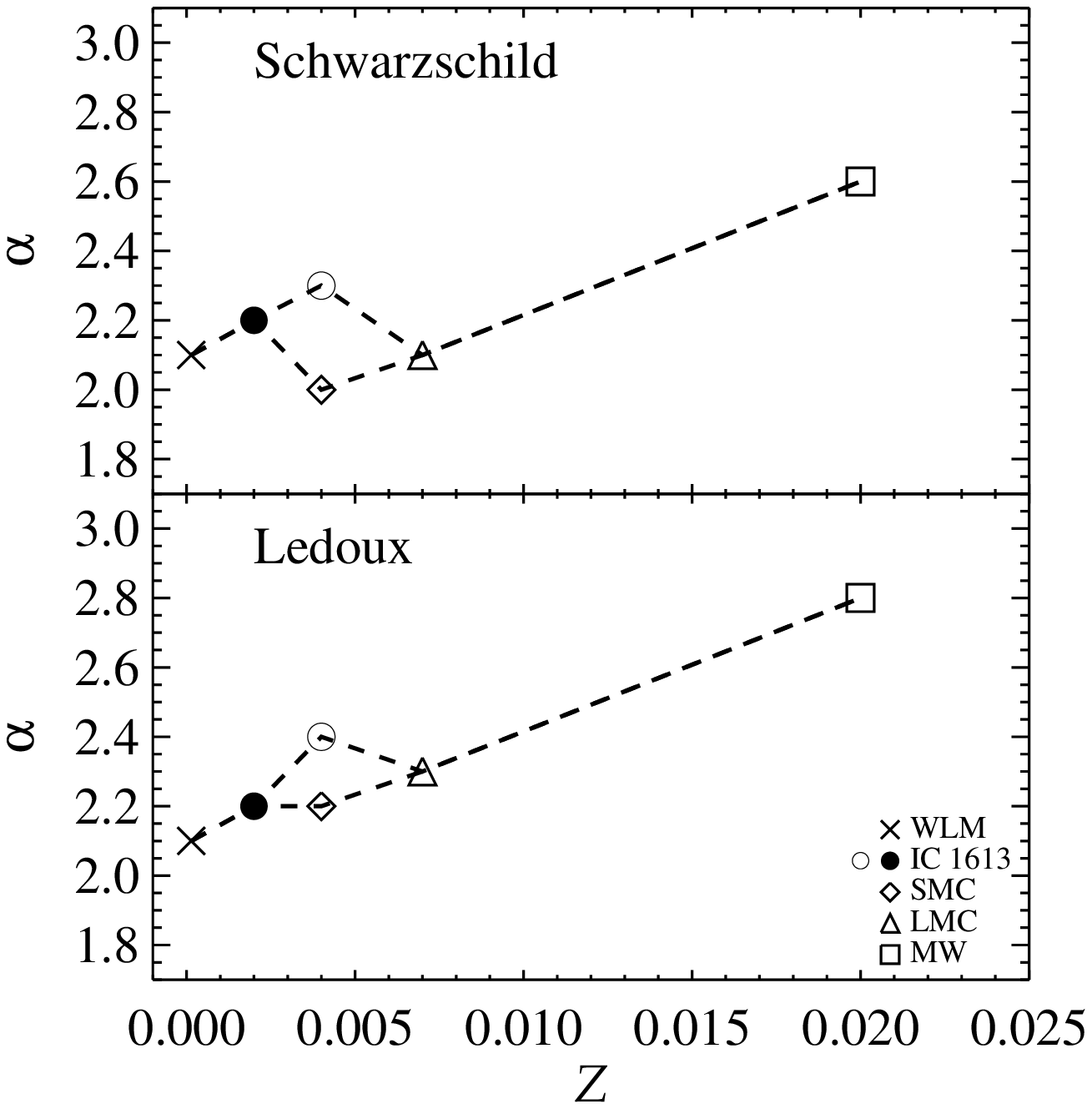}
\caption{The calibrated mixing length values of WLM (Z=0.0014), IC 1613 (Z=0.002 and 0.004), SMC (Z=0.004), LMC (Z=0.007) and the Milky Way (Z=0.02)~\citep{Chun2018} as a function of metallicity obtained with the Schwarzschild (upper panel) and Ledoux (lower panel) models. The mixing length values of SMC, LMC, and
the Milky Way calibrated by $f_\mathrm{ov}=0.15$ models and SED temperatures were adopted.
}
\label{mixing}
\end{figure}

In Figure~\ref{mixing}, we compare the mixing length value of IC 1613 with
those of the SMC (Z=0.004), LMC (Z=0.007), the Milky Way (Z=0.02)
from~\citet{Chun2018}. 
We assume Z=0.002 and Z=0.004 as the metallicity of IC 1613 because the metallicity
of RSGs in this study shows a broad distribution.
The mixing length values of
the different galaxies calibrated by the $f_\mathrm{ov}=0.15$ models and the
SED temperatures in~\citet{Chun2018} were adopted.  
Recently,~\citet{Gonzalez2021} report the effective temperatures
and luminosities of RSGs in Wolf-Lundmark-Mellote (WLM) galaxy by fitting the
SEDs obtained by VLT+XSHOOTER with MARCS model atmospheres. We use their
temperatures and apply the same analysis in this study to obtain the calibrated mixing
length value for WLM, for which we used the metallicity of Z=0.0014 to calculate the
MESA evolution models. We find that evolution models with mixing
length of $\alpha_{\mathrm{MLT}}=2.1$ reproduce well the location of WLM RSGs in HR diagram
for both Schwarzschild and Ledoux convection criteria. The results for WLM
RSGs is included in Figure~\ref{mixing}. 
Here we note that the metallicity and temperature
of IC 1613 without correction discussed in Section 3.1 are very close to those of WLM. 
Thus, the similar mixing length value to WLM is expected. 
If we adopt the metallicity and temperature without correction, we indeed find that $\alpha_{\mathrm{MLT}}=2.1$ value is
the best mixing length for IC 1613.

The metallicity-dependent mixing length trend found by~\citet{Chun2018} becomes
ambiguous by adding the results of IC 1613 and WLM at the low metallicity
regime, in particular for the Schwarzschild case.  The mixing length values
for IC 1613 and WLM are higher than or comparable to that of the SMC in the
Schwarzschild and Ledoux cases, respectively.  The uncertainties in metallicity
and effective temperature of our results seem to make it difficult to find
clear evidence for a metallicity dependent mixing length in the regime of $Z
\le 0.004$.  However, it is evident that the mixing length values for IC 1613
and WLM are still significantly lower than that of the Milky Way.  More
accurate temperature and metallicity measurements from a large sample of RSGs
for several metal poor galaxies are necessary to confirm the
metallicity-dependent mixing length in the metal poor regime.  The metal-poor
([Fe/H]$<-0.8$) star-forming dIrr galaxies in the Local Group such as Pegasus,
Sextans A and Sextans B~\citep[e.g.,][]{Britavskiy2019} would be ideal targets
for future work.

\section{Conclusions and Summary}

We investigate RSGs in the dwarf irregular galaxy IC 1613 in the Local Group using
$J$ band spectra with a low resolution ($R\sim2000$) obtained
by the MMIRS on the MMT telescope.  Among the 72 observed RSG candidates, we
analyze 14 RSGs belonging to IC 1613 of which 3 RSGs were also studied
in previous studies. The effective temperatures and metallicities of the 14
RSGs are derived by synthetic spectral fitting to the observed spectra ranging
from 1.16 $\mu$m to 1.23 $\mu$m where several atomic absorption lines are dominant. 

We find that the average metallicity of RSGs in IC 1613 is [Fe/H]=$-0.69\pm27$
which is consistent with previous results from young massive stars but 
significantly higher than the metallicity of [Fe/H]=$-1.75\sim-1.15$ obtained from old
stellar populations in this galaxy.  We find a broad metallicity distribution with weak double peaks at
[Fe/H]$=-0.4$ and $-0.7$. 
However, we do not 
find the spatial dependence between metal-rich and metal-poor groups. 
On the other hand, the effective temperatures of the RSGs in IC 1613 are systematically 
higher by about 250 K than those of the SMC. Considering the
metallicity-dependent $T_\mathrm{eff}$, the higher $T_\mathrm{eff}$
of IC 1613 might indirectly imply a lower metallicity than the SMC.
However, we do not find a correlation between $T_\mathrm{eff}$ and metallicity among our RSG sample in
IC 1613. By comparing our RSG sample to evolutionary tracks of massive star evolutionary
models in the HR diagram, we find the observed RSGs have masses ranging from
$9M_\sun$ to $23M_\sun$.  
The mixing length value for RSGs in IC 1613 calibrated with evolutionary models is
$\alpha_{\mathrm{MLT}}=2.2-2.4$ for both Schwarzschild and Ledoux convection criteria 
used in the evolutionary models. 

We compared this mixing length value calibrated for IC
1613 with those of SMC (Z=0.004), LMC (Z=0.007) and the Milky Way (Z=0.02)
obtained by \citet{Chun2018}.  We also calibrate the mixing length value as
$\alpha_{\mathrm{MLT}}=2.1$ for the RSGs in WLM (Z=0.0014) by~\citet{Gonzalez2021}.  Although
the trend of decreasing mixing length with decreasing metallicity of host
galaxies weakens at the low metallicity regime of $Z \lesssim 0.004$, it is
evident that the mixing length values for IC 1613 and WLM are lower than that
of the Milky Way.  

Overall, the variations in effective temperatures, mixing length, and spatial
distribution of RSGs by metallicity found in this study need to be further
investigated for a larger sample of RSGs in IC1613, ideally through high
resolution spectroscopic data.  Furthermore, new model atmospheres with a broader range of
temperature and metallicity with alpha-element enhancements are needed to obtain more accurate 
stellar parameters.
An analogous spectroscopic analysis with new model atmospheres for RSGs in several metal-poor dwarf irregular galaxies in the Local
Group will help constrain how the physical properties of RSGs depend on 
metallicity in environments more metal poor than the SMC.

\acknowledgements
We are grateful to Ben Davies, who refereed this paper, for a number of helpful comments.
This work was supported by K-GMT Science Program (PID: MMT-2017B-3) funded
through Korean GMT Project operated by Korea Astronomy and Space Science Institute (KASI).
SHC acknowledges support from the National Research Foundation of Korea (NRF) grant funded
by the Korea government (MSIT) (NRF-2021R1C1C2003511) and the Korea Astronomy and Space 
Science Institute under R\&D program (Project No. 2022-1-830-05) supervised by the Ministry of Science and ICT.
SCY is supported by the National Research Foundation of Korea (NRF) grant (NRF-2019R1A2C2010885).
Observations reported here were obtained at the MMT Observatory, a joint facility of the University of Arizona and the Smithsonian Institution. 

\bibliographystyle{aasjournal}
\bibliography{reference}

\end{document}